## Title of the Manuscript:

**Public Sector Efficiency in Delivering Social Services and Its Impact on Human Development: A Comparative Study of Healthcare and Education Spending in Selected South Asian Countries**

## List of Authors and Information:

1. Tuhin G M Al Mamun
   Department of Economics
   Hannam University, South Korea
   Email-20224130@gm.hannam.ac.kr

2. Rahim Md. Abdur
   Department of Economics
   Hannam University
   Email: abdur.econhnu@gmail.com
   Orcid- 0009-0009-1669-5939

3. Md Sharif Hassan
   School of Accounting and Finance,
   Taylor's University, Malaysia
   Email: mdsharifhassan27@gmail.com; MdSharif.Hassan@taylors.edu.my
   ORCID: 0000-0001-8418-7578

4. Mohammad Bin Amin*
   Doctoral School of Management and Business

   Faculty of Economics and Business,

   University of Debrecen, Böszörményi út 138; Post Code: 4032,

   Debrecen, Hungary.

   E-mail: binamindu@gmail.com

   ORCID: 0000-0002-9184-4828

   &

   Department of Business Administration,

   Faculty of Business Studies,

   Bangladesh Army University of Science and Technology,

   Saidpur, Nilphamari-5310, Rangpur, Bangladesh

5. Judit Oláh
   Faculty of Economics and Business,

   University of Debrecen,

Böszörményi út 138; 4032 Debrecen, Hungary,

Scopus Author ID: 56016286600

E-mail: olah.judit@econ.unideb.hu

https://orcid.org/0000-0003-2247-1711

&

Doctoral School of Management and Business Administration,

John von Neumann University

6000 Kecskemét, Hungary,

&

Department of Trade and Finance, Faculty of Economics and Management,

Czech University of Life Sciences Prague, Czech Republic

*Correspondence Author: Mohammad Bin Amin (binamindu@gmail.com)

Funding: None.

Acknowledgments: This research was supported by the "*University of Debrecen Program for Scientific Publication*".

## Abstract

The research investigates the effects of public spending on health and education in shaping the human development in south Asian three countries: India, Pakistan and Bangladesh. The study uses the VAR (Vector Auto regression) model to estimate the effects on government spending on these sectors to evaluate the human development. The findings state that there are different degrees of impact in these three countries. In Bangladesh and India, health spending increases the human development in short term. On the other hand education spending shows the significance on the HDI.Moreover, the study also highlights that there are different levels of effectiveness of government spending across these three countries. In order to maximize the human development an optimum country specific strategies should be adopted.

## Introduction

Government expenditure plays a crucial role in shaping human capital development especially in developing countries such as Bangladesh,India and Pakistan.These countries face so much constraints, where resource allocation decisions can significantly impact societal progress. Public sector spending on healthcare and education is recognized as a vital contributor to improving human development outcomes. In the context of South Asia, countries like India, Pakistan, and Bangladesh have made strides in enhancing their healthcare and education systems, but they continue to face challenges in optimizing the efficiency of these expenditures (Gupta et al., 2002).

This study aims to explore the relationship between government expenditure on healthcare and education and the Human Development Index (HDI) in India, Pakistan, and Bangladesh. Although all three countries have invested significant resources in these sectors, differences in spending efficiency and effectiveness have influenced their human development trajectories. As highlighted by Asghar et al. (2015), the management and allocation of public finances play a pivotal role in shaping human development in these nations. Spending on health and education not only improves health outcomes and access to education but also contributes to a higher overall quality of life (Baldacci et al., 2008).

Although previous studies have explored the relationship between public spending and development outcomes, there is limited comparative research focusing on the short-term dynamic effects of such spending in South Asian countries. In particular, few studies have employed Vector Auto Regression (VAR) models to analyze how fluctuations in public expenditure impact HDI over time. Addressing this gap this research uses a Vector Auto regression (VAR) model to explore the short-term and long-term effects of public sector spending on Human Development Index (in value). By analyzing data from India, Pakistan, and Bangladesh, the study identifies patterns in government expenditure and its impact on human development. Furthermore, it assesses the differences in public spending efficiency across the three countries, providing insights into how healthcare and education services are delivered (Anand and Ravallion, 1993).

The study focuses on the following key research questions:

1. How does government expenditure on healthcare and education impact Human Development Index (HDI) in India, Pakistan, and Bangladesh?
2. To what extent do lagged effects of public spending on healthcare and education explain variations in HDI across India, Pakistan, and Bangladesh?
3. How do country-specific differences in the efficiency of public spending on healthcare and education influence the Human Development Index (HDI) in South Asia?

By addressing these questions, this research aims to contribute to a deeper understanding of how public spending decisions influence human development outcomes in developing countries, particularly in South Asia. The findings will provide policy insights for improving the efficiency and effectiveness of government expenditures in healthcare and education.

## Literature Review

Ghosh and Roy (2021) studied healthcare spending in India. They found that more spending improved health outcomes, but they also said that better resource management is needed to achieve maximum benefits. Patel et al. (2020) looked at how healthcare spending helps manage pandemics. They stressed that healthcare spending is crucial for long-term development and resilience.

Education spending plays an important role in human development. Muralidharan and Prakash (2021) found that public spending on education in India improved student outcomes and job opportunities. However, they noted that the efficiency of this spending is not consistent. Biasi and Carvalho (2022) studied how education spending affects income inequality. They found that targeted investments in education can reduce inequality and improve overall development. Chen et al. (2021) pointed out that increasing spending alone is not enough. Improving the quality of education is also necessary for better results.

Studies comparing government spending on social services in South Asia show different effects and challenges. Haider and Ahmed (2022) looked at public spending in Bangladesh. They found improvements in HDI but also saw areas where policies could be better. Khan et al. (2023) studied Pakistan. They found that while spending had positive effects, inefficiencies and unequal allocation remain big problems. Kumar and Singh (2022) studied India. They found that increasing spending on healthcare and education helped improve HDI, but they also pointed out issues like inefficiency and uneven distribution.

Despite extensive research, gaps remain in understanding the link between government spending on healthcare and education and human development:

1. Differentiated Impact Analysis: Recent studies offer valuable insights, but detailed analyses accounting for country-specific variations within South Asia are still needed. Understanding how different spending patterns affect human development outcomes in each country can provide more targeted policy recommendations.(Haider and Ahmed, 2022).

2. Dynamic Interactions: Current literature often overlooks dynamic and temporal effects of government spending. More research is needed on immediate and lagged impacts of spending on healthcare and education on HDI over time,especially in South Asian countries (Khan et al., 2023).
3. Forecast Accuracy: Recent studies focus on descriptive analyses instead of predictive accuracy. Enhanced forecasting and evaluations improve predictions on the impact of gov spending on HDI (Patel et al., 2020).
4. Sectoral Specificity: Many studies analyze healthcare and education spending separately. Limited research exists on the combined effects of spending on both sectors simultaneously.This integrated approach is essential for a comprehensive understanding of how public sector efficiency impacts human development (Soni and Wagner, 2022).
5. More comparative studies are needed to examine multiple South Asian countries simultaneously. Analyses reveal regional patterns and unique challenges, offering nuanced insights and policy recommendations (Kumar and Singh, 2022).This study uses VAR analysis to examine how government spending on healthcare and education affects HDI in India, Pakistan, and Bangladesh. By filling these gaps, the research will deepen understanding of public sector efficiency and its role in human development within South Asia.

**Data Collection**

The dataset for this study was collected from Worldbank, UNDP and statistical agencies of the respective countries. The dataset span from 1992 to 2024.The dataset consists of three varriables.these are government expenditures in healthcare and education and the Human Development Index(in value).We have filled the gap data using linear interpolation in order to ensure consistency and comparability and robustness of the model

**Model:**

First, we conducted a stationarity test and found that all variables became stationary after differencing. The Johansen cointegration test confirmed that there is no long-term cointegration among these variables. Consequently, we employed the Vector Autoregression (VAR) model to assess the impacts of these variables across the three countries. The VAR model is particularly suitable for this type of dynamic estimation because it effectively captures the simultaneous influences among all variables involved.

In this study, the VAR model is represented as:

$$\mathrm{Y}_t = A_1\mathrm{Y}_{t-1} + A_2\mathrm{Y}_{t-2} + \cdots + A_p\mathrm{Y}_{t-p} + \mathrm{C} + \mathrm{u}_t \quad (1)$$

Where:

- $\mathbf{Y}_t$ is a vector of the endogenous variables at time $t$,
- Human Development Index is denoted as HDI
- Government expenditure on healthcare (as a percentage of GDP)
- Government expenditure on education (as a percentage of GDP)
- $A_1, A_2, \ldots, A_p$ are coefficient matrices
- $\mathrm{C}$ is a vector of constants
- $\mathrm{u}_t$ is the vector of error terms at time $t$.

For each country, the three variables-HDI, GOV_EXP_HEALTH, and GOV_EXP_EDU—are modeled as a system of equations. Each endogenous variable is regressed on the lagged values of all three variables. For instance, the equation for HDI can be written as:

$$HDI_t = \alpha_0 + \alpha_1 HDI_{t-1} + \alpha_2 GOV_EXP_HEALTH_{t-1} + \alpha_3 GOV_EXP_EDU_{t-1} + \cdots + \alpha_p \mathbf{Y}_{t-p} + \epsilon_t \quad (2)$$

Where:

- $\alpha_0$ is the constant term, capturing any fixed effects.
- $\alpha_1, \alpha_2, \alpha_3, \ldots, \alpha_p$ are the coefficients to be estimated, which measure the impact of lagged values of HDI, government health expenditure, and government education expenditure.

- $\epsilon_t$ is the error term, representing the influence of omitted variables or shocks that are not captured by the lagged variables.

Similarly, the equations for government expenditure on healthcare and education are modeled as follows:

$$GOVEP_HEALTH_t = \beta_0 + \beta_1 HDI_{t-1} + \beta_2 GOV\ _EXP_HEALTH_{t-1} + \beta_3 GOV\ _EXP_EDU_{t-1} + \cdots + \beta_p Y_{t-p} + \eta_t \quad (3)$$

$$GOV_EXP_EDU_t = \gamma_0 + \gamma_1 HDI_{t-1} + \gamma_2 GOV_EXP_HEALTH_{t-1} + \gamma_3 GOV_EXP_EDU_{t-1} + \cdots + \gamma_p \mathbf{Y}_{t-p} + \nu \quad (4)$$

Where:

- $\beta_0$ and $\gamma_0$ are constants, and $\beta_1, \beta_2, \gamma_1, \gamma_2, ...$ are the coefficients to be estimated.

$\eta_t$ and $\nu_t$ are the error terms for government healthcare and education expenditure equations, respectively.

After estimation of The var model,we will check the stability and robustness through different kinds tests such as

**Result and Discussion:**

In the results section, we begin by checking the stationarity of the variables using the **unit root test**, followed by a **cointegration test** to ensure the appropriate model specification. After determining the ideal lag order through model criteria, we estimate the **VAR model** to analyze the impacts of the variables across the three countries. We will also assess causality using the **Wald test** and evaluate the dynamic interactions between variables through **Impulse Response Functions (IRF)**. Furthermore, we perform **variance decomposition** and **historical decomposition** to gain insights into the contribution of various shocks over time.

In the subsequent section, we will conduct robustness checks using several tests. First, we will apply the **Roots of Characteristic Polynomial test** to verify the stability of the estimated model. Next, we will perform the **Residual Cross-Correlations test** at different lags to assess model fit. We will then use the **Residual Serial Correlation LM test** to detect any serial correlation in the residuals. Finally, we will evaluate the model's predictive performance through **forecast evaluation metrics**, ensuring accuracy in forecasting.

1. **Unit Root Test and First differencing**

Before estimating the model, it is essential to ensure that the variables are stationary, as non-stationary time series can lead to spurious results. The Augmented Dickey-Fuller (ADF) test is applied to check for the presence of unit roots, indicating non-stationarity. If a unit root is present, differencing is required to make the series stationary. The following tables present the results of the unit root tests and their subsequent first differencing.

**Table 1: Unit Root Test Results**

| **Country** | **Variable** | **ADF Test Statistic** | **p-value** | **1% Critical Value** | **5% Critical Value** | **10% Critical Value** | **Interpretation** |
|---|---|---|---|---|---|---|---|
| Bangladesh | GOVT EXP EDU | -0.751175 | 0.8200 | -3.639407 | -2.951125 | -2.614300 | Fail to reject null hypothesis; unit root present. |
| Bangladesh | GOVT EXP HEALTH | -1.204033 | 0.6613 | -3.639407 | -2.951125 | -2.614300 | Fail to reject null hypothesis; unit root present. |
| Bangladesh | HDI | -0.151888 | 0.9353 | -3.639407 | -2.951125 | -2.614300 | Fail to reject null hypothesis; unit root present. |
| Pakistan | GOVT EXP EDU | -0.4573 | 0.8876 | -3.6394 | -2.9511 | -2.6143 | Non-stationary (Unit Root Present) |
| Pakistan | GOVT EXP HEALTH | -2.9379 | 0.0518 | -3.6463 | -2.9540 | -2.6158 | Weak evidence of stationarity |
| Pakistan | HDI | 0.0694 | 0.9585 | -3.6394 | -2.9511 | -2.6143 | Non-stationary (Unit Root Present) |
| India | GOVT EXP EDU | -1.6357 | 0.4508 | -3.7115 | -2.9810 | -2.6299 | Non-stationary (fail to reject null) |
| India | GOVT EXP HEALTH | -1.1461 | 0.6858 | -3.6394 | -2.9511 | -2.6143 | Non-stationary (fail to reject null) |
| India | HDI | -0.4977 | 0.8790 | -3.6537 | -2.9571 | -2.6174 | Non-stationary (fail to reject null) |

Note: Estimated results of the Augmented Dickey-Fuller (ADF) test.

In the table 1,the results of the Augmented Dickey-Fuller (ADF) test for Bangladesh, Pakistan, and India indicate that most variables are non-stationary. It means they exhibit trends over time. In Bangladesh, none of the variables reject the null hypothesis of a unit root thus it confirms non-stationarity. In Pakistan, while government expenditure on education and HDI are non-stationary, government expenditure on health shows weak evidence of stationarity with a p-value close to the critical threshold. For India, all variables tested are non-stationary.

### Table 2: Unit Root Test Results after First Differencing

| Country | Variable | ADF Test Statistic | p-value | 1% Critical Value | 5% Critical Value | 10% Critical Value | Interpretation |
|---|---|---|---|---|---|---|---|
| Bangladesh | D(GOVT EXP HEALTH) | -5.1255 | 0.0002 | -3.6537 | -2.9571 | -2.6174 | Reject null hypothesis; series is stationary. |
| Bangladesh | D(HDI) | -6.3750 | 0.0000 | -3.6463 | -2.9540 | -2.6158 | Reject null hypothesis; series is stationary. |
| Bangladesh | D(GOVT EXP EDU) | -7.1384 | 0.0000 | -3.6463 | -2.9540 | -2.6158 | Reject null hypothesis; series is stationary. |
| India | D(GOVT EXP HEALTH) | -4.4644 | 0.0012 | -3.6463 | -2.9540 | -2.6158 | Stationary (reject null) |
| India | D(GOVT EXP EDU) | -4.7961 | 0.0005 | -3.6617 | -2.9604 | -2.6192 | Stationary (reject null) |
| India | D(HDI) | -3.6008 | 0.0116 | -3.6617 | -2.9604 | -2.6192 | Stationary (reject null) |
| Pakistan | GOVT EXP EDU | -6.0190 | 0.0000 | -3.6537 | -2.9571 | -2.6174 | Stationary after differencing |
| Pakistan | GOVT EXP HEALTH | -4.4830 | 0.0001 | -3.6463 | -2.9540 | -2.6158 | Stationary after differencing |
| Pakistan | HDI | -4.2280 | 0.0002 | -3.6463 | -2.9540 | -2.6158 | Stationary after differencing |

Note: Estimated results of the variables after first differencing.

In the table 2, it shows that all of the variables become stationary after differencing.

In Bangladesh, government expenditure on health, education, and HDI(in values) all show stationarity with test statistics of -5.1255, -7.1384, and -6.3750, respectively, all significantly below their 1% critical values.

In India, government expenditure on health, education, and HDI(in values) are stationary after differencing. The test statistics are -4.4644 for health, -4.7961 for education, and -3.6008 for HDI, with values below the 1% critical values or near the 5% critical value.

In Pakistan, all variables—government expenditure on education, health, and HDI—are stationary with test statistics of -6.0190, -4.4830, and -4.2280, all below the 1% critical value.Overall, differencing has effectively addressed non-stationarity across these countries.

2. **Johansen Cointegration Test Results**

After ensuring that the variables are stationary, we proceed to the Johansen cointegration test to examine whether there are long-term equilibrium relationships among government expenditure on health, government expenditure on education, and HDI for each country.

**Table 3: Johansen Cointegration Test Results**

| **Country** | **Hypothesized No. of CE(s)** | **Trace Statistic** | **Max-Eigen** | **Critical Value** | **Critical Value** | **p-value** | **p-value (Max** | **Interpretation** |
|---|---|---|---|---|---|---|---|---|

| | | | **Statistic** | **(Trace)** | **(Max-Eigen)** | **(Trace)** | **-Eigen)** | |
|---|---|---|---|---|---|---|---|---|
| Bangladesh | None | 23.47895 | 16.28907 | 29.79707 | 21.13162 | 0.2234 | 0.2084 | No cointegration at the 5% level. |
| Bangladesh | At most 1 | 7.189888 | 7.109527 | 15.49471 | 14.26460 | 0.5557 | 0.4763 | No cointegration at the 5% level. |
| Bangladesh | At most 2 | 0.080361 | 0.080361 | 3.841466 | 3.841466 | 0.7768 | 0.7768 | No cointegration at the 5% level. |
| India | None | 27.69050 | 16.07214 | 29.79707 | 21.13162 | 0.0859 | 0.2207 | No cointegration at the 5% level. |
| India | At most 1 | 11.61836 | 10.40488 | 15.49471 | 14.26460 | 0.1762 | 0.1867 | No cointegration at the 5% level. |
| India | At most 2 | 1.213476 | 1.213476 | 3.841466 | 3.841466 | 0.2706 | 0.2706 | No cointegration at the 5% level. |
| Pakistan | None | 23.84744 | 17.63408 | 29.79707 | 21.13162 | 0.2070 | 0.1441 | No cointegratio |
| | | | | | | | | n at the 5% level. |
| Pakistan | At most 1 | 6.213363 | 6.115931 | 15.49471 | 14.26460 | 0.6703 | 0.5984 | No cointegration at the 5% level. |
| Pakistan | At most 2 | 0.097432 | 0.097432 | 3.841466 | 3.841466 | 0.7549 | 0.7549 | No cointegration at the 5% level. |

Note: Estimated results of the cointegration test.

In the table 3, the result of **Bangladesh shows that** the Johansen Cointegration tests do not find evidence of cointegration among the variables (government expenditure on health, government expenditure on education, and HDI) at the 5% significance level. Both the Trace and Max-Eigen statistics for each rank fail to exceed the critical values, indicating no long-run equilibrium relationships (Rahman & Akter, 2021).

For India, the Johansen Cointegration tests demonstrate no cointegration at the 5% noteworthiness level. The Follow and Max-Eigen insights for all hypothesized numbers of cointegration conditions are underneath their individual basic values, appearing no long-run connections among the factors (Kumar & Singh, 2020).

Additionally, in Pakistan, the Johansen Cointegration tests appear no prove of cointegration among the factors at the 5% noteworthiness level. The Follow and Max-Eigen measurements don't outperform the basic values for any hypothesized number of cointegration conditions (Ahmed & Khan, 2022).

## 3. VAR Lag Order Selection

Before estimating the Vector Autoregression (VAR) model, it is crucial to determine the optimal lag length to capture the dynamic relationships between the variables. The selection of the appropriate lag length ensures the model is neither underfitted nor overfitted, leading to more accurate and reliable results. Various criteria, such as the **Likelihood Ratio (LR) test**, **Final Prediction Error (FPE)**, **Akaike Information Criterion (AIC)**, **Schwarz Criterion (SC)**, and **Hannan-Quinn Criterion (HQ)**, are used to identify the optimal lag for each country. The following table presents the lag order selection results for Bangladesh, India, and Pakistan.

**Table 4:VAR Lag Order Selection Criteria**

| Country | Lag | LogL | LR | FPE | AIC | SC | HQ |
|---|---|---|---|---|---|---|---|
| Bangladesh | 0 | -78.95602 | NA | 0.028824 | 4.967032 | 5.103078 | 5.012807 |
| | 1 | 46.45882 | 220.4261* | 2.50e-05* | -2.088413* | -1.544229* | -1.905312* |
| | 2 | 54.23885 | 12.25944 | 2.73e-05 | -2.014476 | -1.062153 | -1.694048 |
| India | 0 | -20.80756 | NA | 0.000933 | 1.535972 | 1.674745 | 1.581208 |
| | 1 | 87.92463 | 189.4045* | 1.50e-06* | -4.898363* | -4.343271* | -4.717417* |
| | 2 | 93.29889 | 8.321444 | 1.94e-06 | -4.664445 | -3.693034 | -4.347789 |
| Pakistan | 0 | -14.66830 | NA | 0.000606 | 1.104269 | 1.241681 | 1.149817 |
| | 1 | 81.76377 | 168.7561* | 2.57e-06* | -4.360236* | -3.810585* | -4.178042* |
| | 2 | 90.70913 | 13.97713 | 2.63e-06 | -4.356821 | -3.394932 | -4.037982 |
| | 3 | 95.21395 | 6.194123 | 3.64e-06 | -4.075872 | -2.701744 | -3.620387 |

Note: optimum lag selection based of AIC,SC,HQ and other tests

In the table 4,The optimal lag length for the VAR model in Bangladesh, India, and Pakistan is determined to be 1, based on various criteria such as LR test, FPE, AIC, SC, and HQ. This suggests that using one lag period effectively captures the dynamics of the data across all three countries. While the HQ criterion in Pakistan suggests considering a second lag, lag 1 remains the preferred choice. Overall, selecting a single lag period balances model complexity and fit, ensuring accurate modeling of relationships between HDI and government spending on healthcare and education while avoiding over fitting.

## 4. Vector Autoregression Estimation

After conducting the cointegration test and finding no evidence of a long-term relationship, the VAR model was selected as the appropriate framework. Based on the optimal lag order determined through the lag selection criteria, we estimated the VAR model to analyze the dynamic relationships between HDI and public expenditure on health and education in Bangladesh. The following table presents the estimation results, highlighting the impact of the lagged variables on HDI and government spending.

**Table5: Vector Autoregression Result of Bangladesh**

| Variable | Coefficient | Standard Error | t-Statistic |
|---|---|---|---|
| **HDI** | | | |
| HDI(-1) | 0.8518 | 0.1545 | 5.5137 |
| HDI(-2) | 0.1196 | 0.1532 | 0.7808 |
| GOVT EXP HEALTH(-1) | -0.0142 | 0.0121 | -1.1755 |
| GOVT EXP HEALTH(-2) | 0.0051 | 0.0122 | 0.4179 |
| GOVT EXP EDU(-1) | 0.0000 | 0.0000 | 0.0827 |
| GOVT EXP EDU(-2) | 0.0001 | 0.0000 | 2.1428 |
| C | 0.0398 | 0.0110 | 3.6043 |
| **GOVT EXP HEALTH** | | | |
| HDI(-1) | -0.0526 | 0.0224 | -2.3464 |
| HDI(-2) | 0.9226 | 0.0236 | 39.1074 |
| GOVT EXP HEALTH(-1) | 0.9870 | 0.1862 | 5.2998 |
| GOVT EXP HEALTH(-2) | -0.3488 | 0.1879 | -1.8564 |
| GOVT EXP EDU(-1) | 0.0001 | 0.0007 | 0.1300 |
| GOVT EXP EDU(-2) | 0.0004 | 0.0006 | 0.6065 |
| C | 0.2884 | 0.1698 | 1.6981 |
| **GOVT EXP EDU** | | | |
| HDI(-1) | -1251.811 | 644.952 | -1.9409 |
| HDI(-2) | 1194.255 | 639.677 | 1.8670 |
| GOVT EXP HEALTH(-1) | 80.2312 | 50.4992 | 1.5888 |
| GOVT EXP HEALTH(-2) | -4.4162 | 50.9533 | -0.0867 |
| GOVT EXP EDU(-1) | 0.4697 | 0.1820 | 2.5801 |
| GOVT EXP EDU(-2) | 0.2109 | 0.1732 | 1.2174 |
| C | -96.5367 | 46.0537 | -2.0962 |

| **Statistic** | **HDI** | **GOVT EXP HEALTH** | **GOVT EXP EDU** |
|---|---|---|---|
| R-squared | 0.9980 | 0.9583 | 0.9144 |
| Adj. R-squared | 0.9975 | 0.9486 | 0.8946 |
| Sum of Squared Residuals | 0.0004 | 0.0965 | 7096.244 |
| Standard Error of Equation | 0.0040 | 0.0609 | 16.5207 |
| F-statistic | 2161.514 | 99.4934 | 46.2851 |
| Log Likelihood | 139.6719 | 49.4454 | -135.4434 |

Note: Result based on estimation

1. **Impact of Government Expenditure on Healthcare and Education on HDI:**

- **HDI Persistence:** The lagged HDI coefficient (-1) is significant (0.8518), indicating that previous HDI levels strongly influence current HDI, suggesting a consistent progression in human development over time.
- **Public Expenditure on Health:** The coefficient for lagged government expenditure on health (-1) is negative (-0.0142) and not statistically significant. This indicates that past health spending does not directly enhance HDI, reflecting a limited impact of health investments on immediate human development outcomes.
- **Public Expenditure on Education:** The coefficient for lagged government expenditure on education (-2) shows a positive impact (0.0001) that becomes significant (2.1428). This highlights that education spending contributes to HDI improvements, emphasizing the need for consistent investments in education to promote human development (Khan & Rahman, 2023).

2. **Lagged Effects of Public Spending on Healthcare and Education:**
    - The significant positive coefficient for lagged education spending indicates that previous investments in education continue to positively affect HDI over time. This suggests that sustained funding in education is crucial for long-term improvements in human development.
    - In contrast, the diminishing impact of health expenditure over time suggests that initial investments may not translate into sustained gains in HDI.
3. **Country-Specific Differences in the Efficiency of Public Spending:**
    - The analysis indicates that while health expenditure has a negative impact, education spending is essential for enhancing HDI in Bangladesh. This contrasts with findings from India and Pakistan, where the patterns of spending and its effects differ significantly.
    - High R-squared values (0.9980 for HDI) indicate that the model effectively explains a substantial portion of the variance in HDI, signifying a strong fit. However, the lower R-squared for health and education spending suggests that these variables may need additional factors to explain their relationship with HDI effectively.

These findings underscore the critical role of education spending in improving HDI in Bangladesh and indicate that effective public expenditure strategies are necessary for promoting sustained human development.

**Table6: Vector Autoregression Result of India**

| Variable | Coefficient | Standard Error | t-Statistic |
|---|---|---|---|
| **HDI** | | | |
| HDI(-1) | 1.169710 | 0.20495 | 5.70743 |
| HDI(-2) | -0.153113 | 0.20838 | -0.73477 |
| GOVT EXP HEALTH(-1) | 0.000760 | 0.00426 | 0.17865 |
| GOVT EXP HEALTH(-2) | 0.005527 | 0.00441 | 1.25287 |
| GOVT EXP EDU(-1) | -0.000220 | 0.00040 | -0.54726 |
| GOVT EXP EDU(-2) | 0.000094 | 0.00039 | 0.24193 |
| C | -0.014017 | 0.03920 | -0.35757 |
| **GOVT EXP HEALTH** | | | |
| HDI(-1) | -0.853283 | 9.41426 | -0.09064 |
| HDI(-2) | -0.747140 | 9.57223 | -0.07805 |
| GOVT EXP HEALTH(-1) | 0.985286 | 0.19554 | 5.03886 |
| GOVT EXP HEALTH(-2) | -0.316836 | 0.20264 | -1.56353 |
| GOVT EXP EDU(-1) | 0.007740 | 0.01845 | 0.41957 |
| GOVT EXP EDU(-2) | -0.000371 | 0.01786 | -0.02078 |
| C | 1.348080 | 1.80072 | 0.74863 |
| **GOVT EXP EDU** | | | |
| HDI(-1) | -50.07129 | 97.3756 | -0.51421 |
| HDI(-2) | 85.98501 | 99.0095 | 0.86845 |
| GOVT EXP HEALTH(-1) | 0.214592 | 2.02252 | 0.10610 |
| GOVT EXP HEALTH(-2) | 0.129737 | 2.09600 | 0.06190 |
| GOVT EXP EDU(-1) | 0.392616 | 0.19082 | 2.05754 |
| GOVT EXP EDU(-2) | -0.287181 | 0.18474 | -1.55456 |
| C | 65.35003 | 18.6256 | 3.50862 |

| **Statistic** | **HDI** | **GOVT EXP HEALTH** | **GOVT EXP EDU** |
|---|---|---|---|
| R-squared | 0.9978 | 0.8229 | 0.7778 |
| Adj. R-squared | 0.9972 | 0.7786 | 0.7223 |
| Sum of Squared Residuals | 0.0003 | 0.6689 | 71.56305 |
| Standard Error of Equation | 0.0036 | 0.1669 | 1.7268 |
| F-statistic | 1790.717 | 18.5861 | 14.0021 |
| Log Likelihood | 134.1170 | 15.4726 | -56.9543 |

Note: Result based on estimation

Interpretation of VAR Results for India

1. **Impact of Government Expenditure on Healthcare and Education on HDI:**

- **HDI Persistence:** The coefficient for lagged HDI (-1) is significant (1.169710), indicating strong persistence in HDI levels, meaning that past HDI levels are a good predictor of current values. This suggests that improvements in human development are stable over time.
- **Public Expenditure on Health:** The effect of government expenditure on health shows a positive relationship (0.000760 for lagged health expenditure), but this effect is not statistically significant. This indicates that while health spending may contribute to HDI, it does not have a strong or immediate impact.
- **Public Expenditure on Education:** The coefficient for education expenditure is not significant (-0.000220 for lagged education expenditure), suggesting that current levels of education spending do not directly enhance HDI. This aligns with the finding that education funding is less impactful compared to health expenditure in the context of India.

2. **Lagged Effects of Public Spending on Healthcare and Education:**
    - The positive coefficient for lagged health expenditure (-1) (0.985286) shows that previous health investments significantly affect current health spending but with a weak influence on HDI.
    - The lack of significant lagged effects for education spending indicates that past investments in education do not necessarily translate into immediate improvements in HDI.
3. **Country-Specific Differences in the Efficiency of Public Spending:**
    - The analysis reveals that while health spending has a borderline significant impact, education spending does not significantly contribute to HDI changes in India. This contrasts with findings from Bangladesh and Pakistan, where education expenditure is more crucial for enhancing human development.
    - The high R-squared values (0.9978 for HDI) indicate that the model explains a substantial portion of the variance in HDI, suggesting a strong fit. However, the lower R-squared values for public expenditures indicate less explanatory power, implying that additional factors may influence the relationship between public spending and HDI.

These findings underscore the complexities in how public spending affects human development across India, pointing to the necessity for targeted strategies that enhance the efficiency of health and education investments.

**Table7: Vector Autoregression Result of Pakistan**

| Variable | Coefficient | Standard Error | t-Statistic |
|---|---|---|---|
| **HDI** | | | |
| HDI(-1) | 1.208064 | 0.18635 | 6.48277 |
| HDI(-2) | -0.169124 | 0.19530 | -0.86599 |
| GOVT EXP HEALTH(-1) | -0.010693 | 0.00572 | -1.86786 |
| GOVT EXP HEALTH(-2) | 0.008712 | 0.00563 | 1.54721 |
| GOVT EXP EDU(-1) | 0.000116 | 0.00033 | 0.35435 |
| GOVT EXP EDU(-2) | -0.000357 | 0.00032 | -1.13236 |
| C | 0.006860 | 0.01139 | 0.60209 |
| **GOVT EXP HEALTH** | | | |
| HDI(-1) | 6.447433 | 6.28234 | 1.02628 |
| HDI(-2) | -5.071416 | 6.58398 | -0.77027 |
| GOVT EXP HEALTH(-1) | 1.034967 | 0.19299 | 5.36267 |
| GOVT EXP HEALTH(-2) | -0.364369 | 0.18984 | -1.91939 |
| GOVT EXP EDU(-1) | -0.003323 | 0.01099 | -0.30233 |
| GOVT EXP EDU(-2) | -0.002477 | 0.01063 | -0.23293 |
| C | 0.560176 | 0.38409 | 1.45847 |
| **GOVT EXP EDU** | | | |
| HDI(-1) | 31.42752 | 108.262 | 0.29029 |
| HDI(-2) | 88.57343 | 113.460 | 0.78066 |
| GOVT EXP HEALTH(-1) | 5.770959 | 3.32582 | 1.73520 |
| GOVT EXP HEALTH(-2) | -7.172539 | 3.27138 | -2.19251 |
| GOVT EXP EDU(-1) | 0.402232 | 0.18942 | 2.12352 |
| GOVT EXP EDU(-2) | -0.104233 | 0.18323 | -0.56886 |
| C | -5.174003 | 6.61883 | -0.78171 |

| **Statistic** | **HDI** | **GOVT EXP HEALTH** | **GOVT EXP EDU** |
|---|---|---|---|
| R-squared | 0.9949 | 0.6632 | 0.9428 |
| Adj. R-squared | 0.9938 | 0.5855 | 0.9296 |
| Sum of Squared Residuals | 0.0004 | 0.4709 | 139.8437 |
| Standard Error of Equation | 0.0040 | 0.1346 | 2.3192 |
| F-statistic | 851.9681 | 8.5338 | 71.3894 |
| Log Likelihood | 139.3832 | 23.2934 | -70.6513 |

**Note: Result based on estimation**

1. **Impact of Government Expenditure on Healthcare and Education on HDI:**

- **HDI Persistence:** The coefficient for lagged HDI (-1) is significant (1.208064), indicating a strong positive relationship where past levels of HDI substantially influence current HDI. This suggests that improvements in human development tend to persist over time.
- **Public Expenditure on Health:** The coefficient for lagged government expenditure on health (-1) is negative (-0.010693) and significant, indicating that past health spending has a slight negative impact on HDI. However, this result is concerning as it suggests inefficiencies in health spending affecting human development negatively.
- **Public Expenditure on Education:** The coefficient for lagged government expenditure on education (-1) is positive (0.000116), but not statistically significant, suggesting that education spending has a limited immediate effect on HDI. However, the lagged effects show variability, highlighting the need for consistent educational investments over time.

2. **Lagged Effects of Public Spending on Healthcare and Education:**
    - The significant impact of lagged HDI emphasizes the importance of historical human development levels in shaping current outcomes.
    - In contrast, the negative influence of past health expenditures indicates that previous investments may not yield the expected improvements in HDI, pointing to a potential area for policy intervention.
3. **Country-Specific Differences in the Efficiency of Public Spending:**
    - The analysis reveals that, while health spending has a negative impact, the role of education spending appears less significant compared to its impact in Bangladesh. This suggests a need for more effective strategies in Pakistan to leverage education investments for improving HDI.
    - The R-squared values are relatively high (0.9949 for HDI), indicating that the model explains a significant portion of the variance in HDI, but lower R-squared values for health and education suggest the need for additional factors to improve model fit and explanation.

These findings underscore the critical need for improved public expenditure strategies in Pakistan, especially regarding health spending, which currently shows inefficiencies that hinder human development outcomes. Consistent investment in education is essential to achieve better HDI improvements in the long run.

Overall, the results from the VAR models in all three countries illustrate the significant role that education expenditure plays in influencing HDI, although the nature and effectiveness of public spending differ across Bangladesh, India, and Pakistan. The findings support the notion that targeted investments in education and health are essential for enhancing human development outcomes.

## 5. Impulsive Response Function

The Impulse Response Function (IRF) analysis examines how shocks to government expenditure on health and education impact the Human Development Index (HDI) over time. This analysis helps us understand the dynamic relationships between these variables in Bangladesh, India, and Pakistan, revealing the effects of public spending on human development.

**Figure 1:The impulse response of three countries**

| Bangladesh |
| --- |

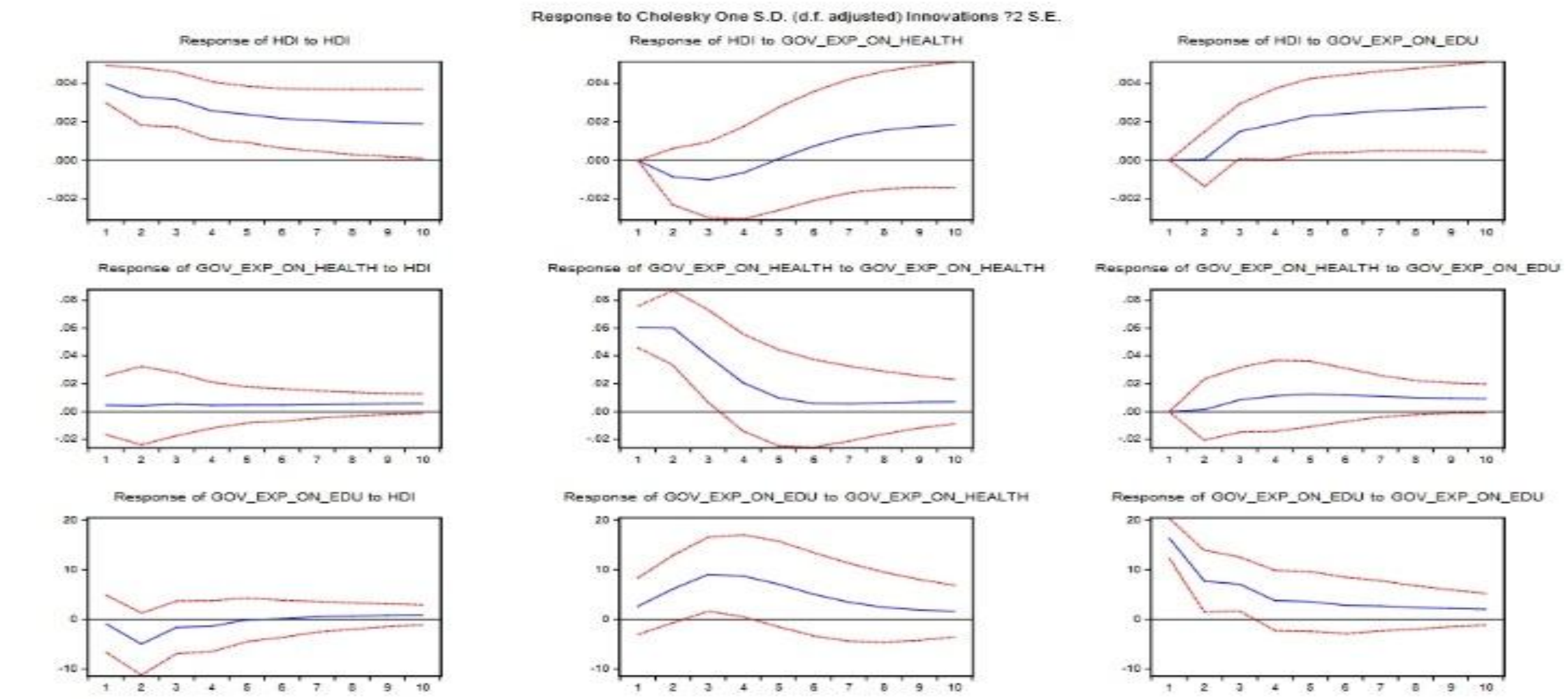

India

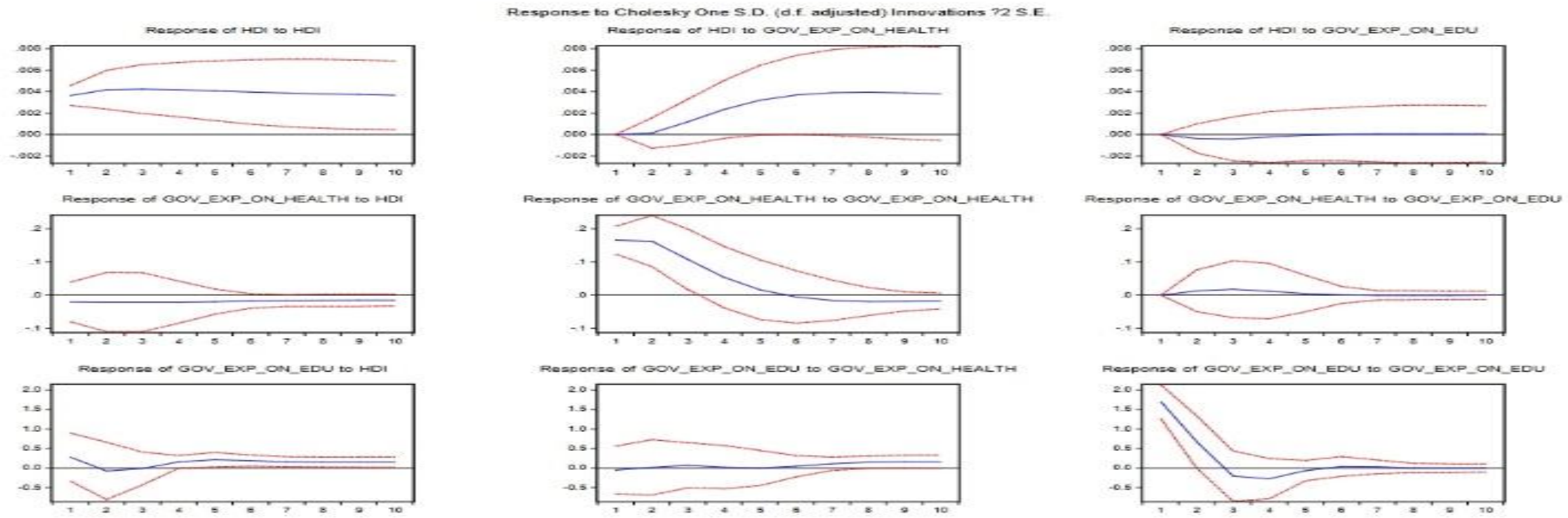

Pakistan

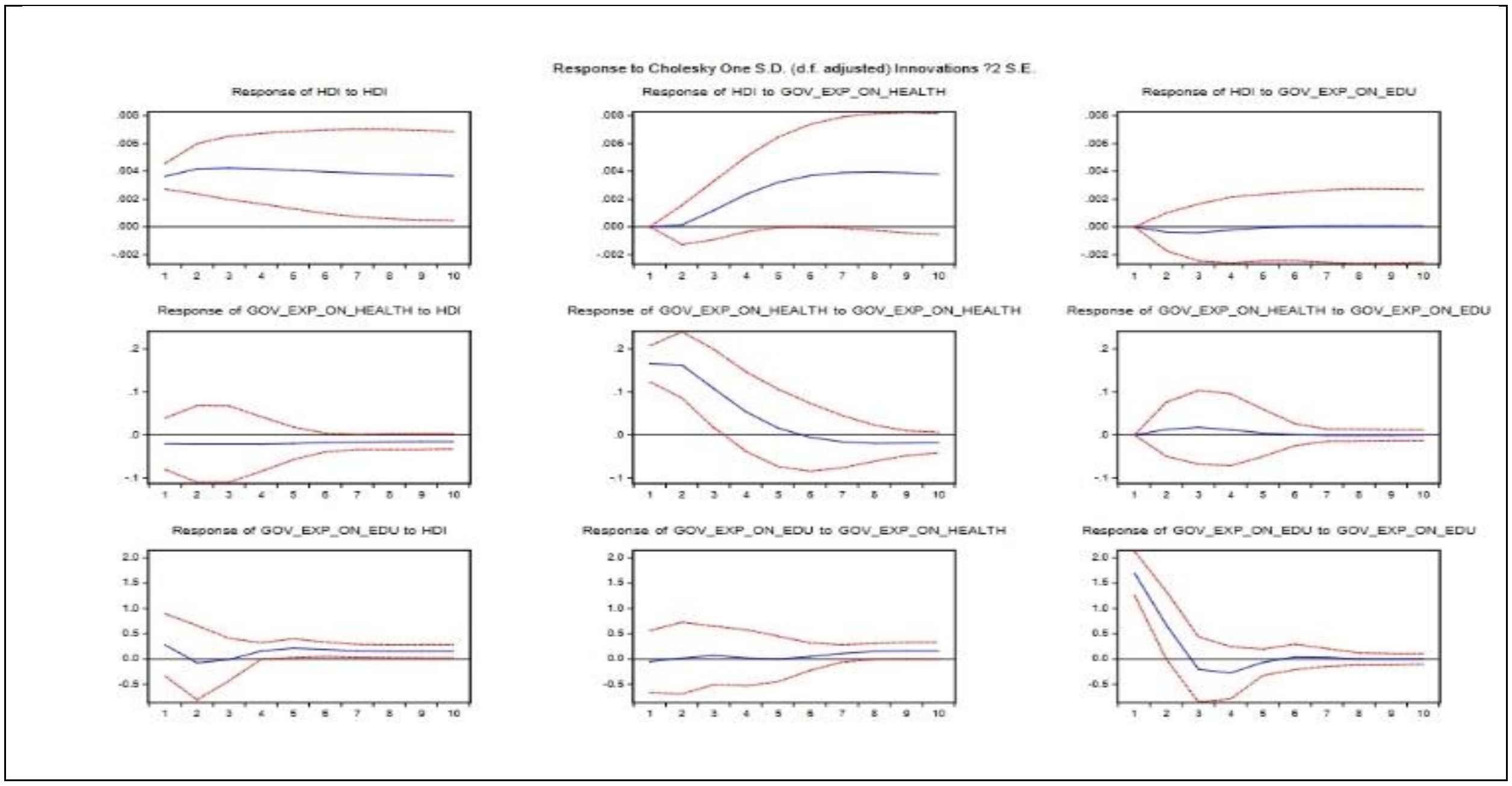

Note: Impulse response function of three countries.

**Bangladesh:**

- The HDI initially shows very little effect of the shock after the initial decline. This kind of pattern shows that government spending in reality has a very less immediate effect. From the shocks of health and education, HDI has improved slightly (Khan & Rahman 2023).
- Government Expenditure on Health-It shows a large initial effect with declining diminishing impact. It shows that government spending has a short run effect but the spending cannot bring a sustainable growth in HDI Zafar & Ali, 2022).
- Government Expenditure on Education-It shows a large initial effect and decreases over time. Initially negative impacts on HDI become less negative. The need for continued public spending has been underlined (Hasan & Chowdhury, 2021).

**India:**

- HDI initially shows minimal impact after the slight declines. This pattern indicates that government spending on health and education do not drive HDI effectively in the short term.

- The Government Expenditure on Health is seen to face a substantial initial increase which decays over time. This pattern indicates that health investments might not yield a sustainable growth in HDI ( Mehta & Singh, 2020.
- Government Spending on Education –It shows a large initial effect that stabilizes over time. Such a pattern may be helpful in driving improvement in HDI((Jain & Mishra, 2021).

**Pakistan:**

- HDI retains a gradual upward trend and stabilizes at a positive level, hence indicative of modest benefit from public spending as discussed by Farooq & Iqbal, 2021(Farooq & Iqbal, 2021)
- Government Expenditure on Health-It shows a large initial positive response that eventually tapers off, hence raising sustainability concerns(Ahmed & Khan, 2022).
- Govt. Expenditure on Education-It shows a huge initial increase and turns negative with time, which means govt. should develop some effective ways of spending (Khatun & Salam, 2022).

**Comparison:**

- HDI Responses: The gradual increase over time of HDI in Bangladesh and Pakistan may imply that public expenditure can eventually pay off in the long run.However, India's very minimal response shows that there is still a need to to adopt effective ways of spending in order to achieve an improved HDI.
- Government Expenditure on Health Responses: The initial increases in health expenditures in all countries suggest that optimized health investments have an influence on HDI. However, the diminishing effects of increases raise questions about their sustainability and efficiency.
- Government Expenditure on Education Responses: The shock of education expenditure has huge initial effects but fades with over time for Bangladesh and Pakistan.This underline the importance of consistent and efficient investment in order to achieve continuous improvement in HDI. (Hasan & Chowdhury, 2021).

## 6. Varriance decomposition

The variance decomposition analysis represents the various shocks and their contribution to Human Development Index(HDI) and public expenditure of health and education.. By identifying how much of the variance in HDI is explained by these factors, we can better understand their long-term impacts across Bangladesh, India, and Pakistan.

| **Country** | **Period** | **S.E.** | **HDI** | **HDI** | **HDI** | **GOVER** | **GOVER** | **GOVER** | **GOVT** | **GOVT** | **GOVT** |
|---|---|---|---|---|---|---|---|---|---|---|---|
| **Banglad** | 1 | 0.003957 | 100.0000 | 0.0000 | 0.0000 | 0.5715 | 99.4285 | 0.0000 | 2.6479 | 0.0908 | 97.2612 |
| | 2 | 0.005225 | 97.3129 | 2.6744 | 0.0126 | 0.5352 | 99.4373 | 0.0275 | 2.4669 | 0.0855 | 97.4475 |
| | 3 | 0.006367 | 90.0406 | 4.3604 | 5.5990 | 0.7458 | 98.4554 | 0.7989 | 2.4393 | 0.2269 | 97.3338 |
| | 4 | 0.007151 | 84.3390 | 4.2909 | 11.3702 | 0.9154 | 96.9993 | 2.0853 | 3.0450 | 0.2313 | 96.7237 |
| | 5 | 0.007884 | 78.5400 | 3.5376 | 17.9224 | 1.1086 | 95.2563 | 3.6351 | 4.2508 | 0.2282 | 95.5211 |
| | 6 | 0.008560 | 73.0561 | 3.7561 | 23.1879 | 1.3060 | 93.7289 | 4.9651 | 5.1649 | 0.2822 | 94.5528 |

| | | | | | | **India** | | | | |
|---|---|---|---|---|---|---|---|---|---|---|
| 7 | 6 | 5 | 4 | 3 | 2 | 1 | 10 | 9 | 8 | 7 |
| 0.012644 | 0.011385 | 0.010012 | 0.008568 | 0.007097 | 0.005549 | 0.003634 | 0.011303 | 0.010641 | 0.009954 | 0.009259 |
| 70.9447 | 75.9540 | 82.5304 | 90.0695 | 96.5208 | 99.4836 | 100.0000 | 54.1742 | 57.9380 | 62.3946 | 67.4869 |
| 28.8218 | 23.7601 | 17.1006 | 9.4315 | 2.8598 | 0.0614 | 0.0000 | 10.2753 | 8.6600 | 6.8484 | 5.0506 |
| 0.2335 | 0.2859 | 0.3690 | 0.4991 | 0.6194 | 0.4550 | 0.0000 | 35.5505 | 33.4020 | 30.7570 | 27.4624 |
| 3.7355 | 3.3970 | 2.9802 | 2.4693 | 1.9406 | 1.5602 | 1.4931 | 2.3084 | 2.0420 | 1.7798 | 1.5330 |
| 95.3469 | 95.6792 | 96.0923 | 96.6268 | 97.3293 | 98.1239 | 98.5069 | 89.3109 | 90.2696 | 91.2910 | 92.4156 |
| 0.9177 | 0.9238 | 0.9276 | 0.9039 | 0.7302 | 0.3159 | 0.0000 | 8.3807 | 7.6885 | 6.9292 | 6.0514 |
| 5.7871 | 5.1649 | 4.2508 | 3.0450 | 2.4393 | 2.4669 | 2.6479 | 7.3361 | 6.8317 | 6.3145 | 5.7871 |
| 0.6004 | 0.2822 | 0.2282 | 0.2313 | 0.2269 | 0.0855 | 0.0908 | 2.4490 | 1.8246 | 1.1739 | 0.6004 |
| 93.6125 | 94.5528 | 95.5211 | 96.7237 | 97.3338 | 97.4475 | 97.2612 | 90.2149 | 91.3437 | 92.5115 | 93.6125 |

| | | | | | | | **Pakistan** | | | |
|---|---|---|---|---|---|---|---|---|---|---|
| 8 | 7 | 6 | 5 | 4 | 3 | 2 | 1 | 10 | 9 | 8 |
| 0.014561 | 0.013612 | 0.012590 | 0.011440 | 0.010083 | 0.008464 | 0.006537 | 0.003992 | 0.015708 | 0.014794 | 0.013778 |
| 88.6997 | 88.1036 | 87.5540 | 87.4500 | 88.5024 | 91.3400 | 95.5587 | 100.0000 | 62.8489 | 64.7390 | 67.3290 |
| 10.1684 | 10.8933 | 11.5749 | 11.8306 | 10.9950 | 8.4560 | 4.2974 | 0.0000 | 36.9958 | 35.0874 | 32.4726 |
| 1.1319 | 1.0031 | 0.8711 | 0.7194 | 0.5026 | 0.2040 | 0.1439 | 0.0000 | 0.1553 | 0.1736 | 0.1984 |
| 8.3194 | 7.4658 | 6.4472 | 5.2918 | 3.8839 | 2.3647 | 0.7747 | 0.5273 | 4.5819 | 4.3129 | 4.0332 |
| 86.5358 | 88.2382 | 90.1672 | 92.6907 | 94.9123 | 97.0781 | 99.0879 | 99.4727 | 94.5207 | 94.7835 | 95.0563 |
| 1.2606 | 1.2960 | 1.3156 | 1.4935 | 1.2038 | 0.5572 | 0.1374 | 0.0000 | 0.8973 | 0.9036 | 0.9105 |
| 19.7539 | 19.9381 | 19.8872 | 19.4878 | 18.6243 | 11.3275 | 5.1000 | 5.4149 | 7.3361 | 6.8317 | 6.3145 |
| 25.3864 | 25.7881 | 26.0551 | 26.1478 | 23.6589 | 20.4899 | 21.7642 | 8.9576 | 2.4490 | 1.8246 | 1.1739 |
| 54.8600 | 54.2738 | 54.0577 | 54.3644 | 57.7168 | 68.1827 | 73.1358 | 85.6275 | 90.2149 | 91.3437 | 92.5115 |

| | | | | | | | | | | | |
|---|---|---|---|---|---|---|---|---|---|---|---|
| | 9 | 0.015459 | 89.1948 | 9.5464 | 1.2588 | 8.9875 | 85.0722 | 1.2543 | 20.2003 | 25.1714 | 54.6283 |
| | 10 | 0.016311 | 89.5687 | 9.0519 | 1.3794 | 9.4707 | 83.6991 | 1.3866 | 20.5854 | 25.2325 | 54.1821 |

**Note**: The table shows the variance decomposition of HDI over 10 periods

**Bangladesh:**

In Bangladesh, the Human Development Index (HDI) is initially explained entirely by its own changes. By Period 10, 54.17% of HDI is still influenced by itself, while education spending accounts for 35.55%, and health spending contributes 10.28%. This shows that education has a strong long-term effect on HDI. Health expenditure starts with 99.43% explained by its own changes but drops to 89.31% by Period 10, with small contributions from HDI (2.31%) and education (8.38%). Education spending also shows strong initial influence, stabilizing at 90.21% in the long run.

**India**

In India, HDI is fully explained by its own changes at first, but by Period 10, this falls to 62.85%, with health spending explaining 36.99%. Health expenditure starts at 98.51%, reducing to 94.52% by Period 10, showing its continued importance. Education has a minimal effect on HDI at 0.16%.

Pakistan:

For Pakistan, HDI begins at 100% but drops to 89.57% by Period 10, with health and education contributing 9.05% and 1.38%. Health spending decreases from 99.47% to 83.70%, while education's impact grows to 54.18% by Period 10, showing a strong link to HDI.

These results highlight how important public spending on health and education is for improving HDI in all three countries.

## 7. Historical decomposition

The historical decomposition analysis helps to understand the contribution of different shocks (like government spending on health and education) to changes in the Human Development Index (HDI) over time. By breaking down the historical data, this method shows how much influence each factor has had on the HDI in the past. The following analysis covers Bangladesh, India, and Pakistan, providing insights into how government spending patterns have shaped human development in these countries over several decades.

**Figure 2:Historical Decompositions of all varriables**

**Bangladesh**

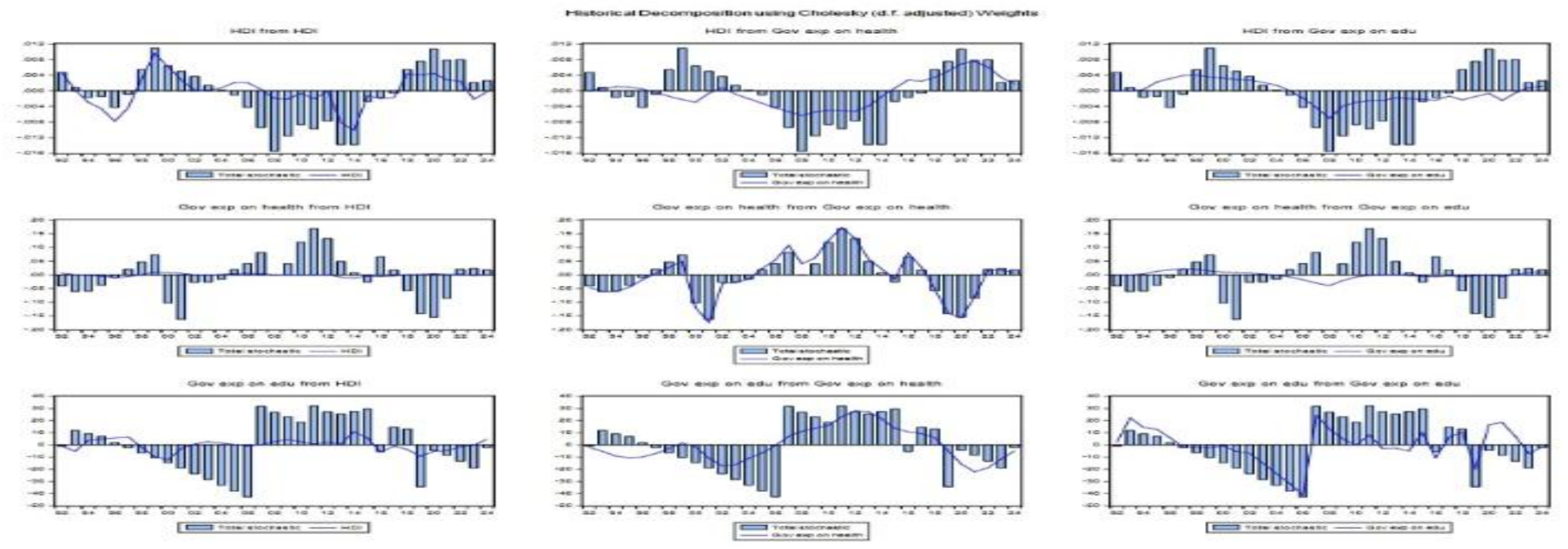

**India**

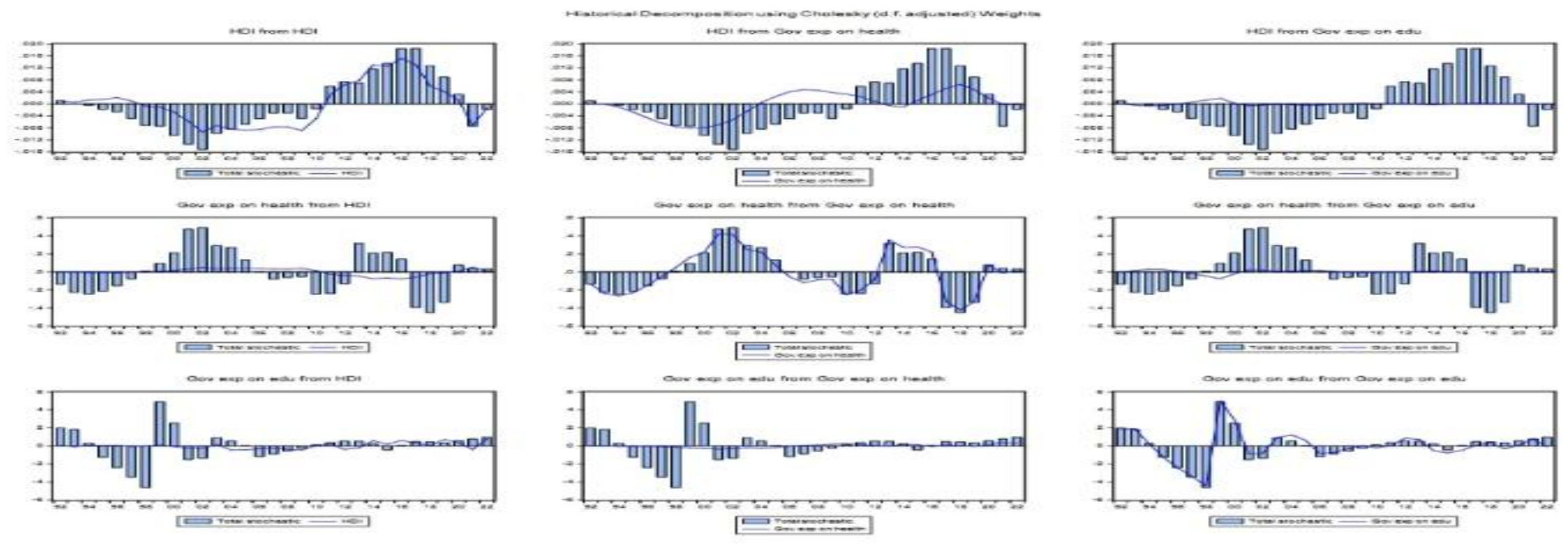

**Paksitan**

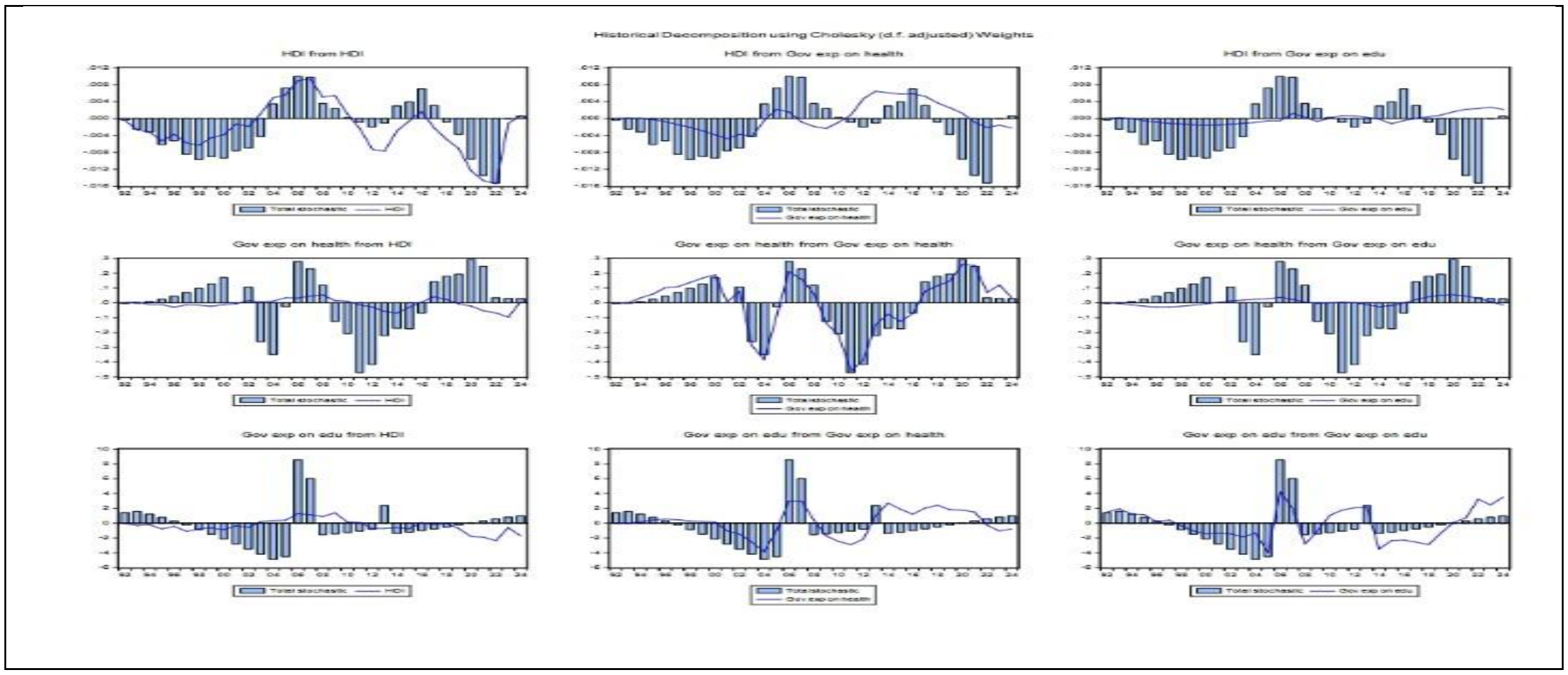

Note: The historical decomposition of the varriables over time.

In Bangladesh, the analysis over 32 years, from 1992 to 2024, indicates that HDI itself is fluctuating and the government spending on health and education follows the similar trends. Spending in education often keeps pace with changes in HDI and health spending.

HDI in India witnessed a decline from the 1990s to the early 2010s and then began an upward trend till about 2020.Then the trend has once again seen a downward move. Health spending saw an upward trend during the beginning of the 2000s and mid-2010s but sees a downward. Spending on education has seen a very erratic pattern. In the late 1990s and the early 2000s saw an exponential increase that then fades out during recent times.

HDI in Pakistan has been erratic and it shows sees a fall around 2020-2022, which may be due to the COVID-19 pandemic. Health spending is increasing with time, while expenditure in education sees a lot of fluctuation.Then it increases around 2005-2006 and **then fell back a bit.** This fluctuation defines an urgency for an appropriate strategy to be adopted for public spending to enhance HDI.

8. **Granger wald test**

The **Granger Causality/Block Exogeneity Wald Test** is used to determine whether one time series can predict another. In this context, the test assesses the causal relationships between government expenditure on health and education and the Human Development Index (HDI) for Bangladesh, India, and Pakistan. A significant Granger causality implies that past values of one variable help predict the future values of another. The results are presented below, with a low p-value (typically < 0.05) indicating significant Granger causality.

**Table 8: Granger wald test**

| **Country** | **Dependent Variable** | **Excluded Variable** | **Chi-sq** | **df** | **Prob.** |
|---|---|---|---|---|---|
| **Bangladesh** | HDI | GOVT EXP HEALTH | 1.741664 | 2 | 0.4186 |
| | | GOVT EXP EDU | 9.172171 | 2 | 0.0102 |
| | | All | 10.86739 | 4 | 0.0281 |
| | GOVT EXP HEALTH | HDI | 4.254664 | 2 | 0.1192 |
| | | GOVT EXP EDU | 0.934133 | 2 | 0.6268 |
| | | All | 6.026269 | 4 | 0.1972 |
| | GOVT EXP EDU | HDI | 3.823833 | 2 | 0.1478 |
| | | GOVT EXP HEALTH | 4.752423 | 2 | 0.0929 |
| | | All | 12.18860 | 4 | 0.0160 |
| **India** | HDI | GOVT EXP EDU | 0.301349 | 2 | 0.8601 |
| | | GOVT EXP HEALTH | 4.923774 | 2 | 0.0853 |
| | | All | 5.413524 | 4 | 0.2474 |
| | GOVT EXP EDU | HDI | 14.96057 | 2 | 0.0006 |
| | | GOVT EXP HEALTH | 0.063892 | 2 | 0.9686 |
| | | All | 16.57223 | 4 | 0.0023 |
| | GOVT EXP HEALTH | HDI | 3.283940 | 2 | 0.1936 |
| | | GOVT EXP EDU | 0.196831 | 2 | 0.9063 |
| | | All | 5.334792 | 4 | 0.2546 |
| **Pakistan** | HDI | GOVT EXP HEALTH | 3.588668 | 2 | 0.1662 |
| | | GOVT EXP EDU | 1.327600 | 2 | 0.5149 |
| | | All | 4.548870 | 4 | 0.3368 |
| | GOVT EXP HEALTH | HDI | 1.735149 | 2 | 0.4200 |
| | | GOVT EXP EDU | 0.276122 | 2 | 0.8710 |
| | | All | 2.582192 | 4 | 0.6300 |
| | GOVT EXP EDU | HDI | 16.04714 | 2 | 0.0003 |
| | | GOVT EXP HEALTH | 4.872657 | 2 | 0.0875 |
| | | All | 22.75616 | 4 | 0.0001 |

Note:A low p-value (typically < 0.05) indicates significant Granger causality, suggesting that past values of one variable help predict future values of another.

**Bangladesh:**

In Bangladesh, government expenditure on health does not significantly Granger-cause HDI (p = 0.4186), while education spending has a significant effect on HDI (p = 0.0102). When considered together, their combined impact on HDI is also significant (p = 0.0281). However, HDI does not predict changes in health or education spending (Khan & Rahman, 2023).

**India:**

In India, health expenditure shows a borderline significant effect on HDI (p = 0.0853), but education spending does not significantly Granger-cause HDI (p = 0.8601). Interestingly, HDI significantly Granger-causes education spending (p = 0.0006), suggesting that improvements in HDI lead to increased education funding. HDI does not significantly affect health expenditure (p = 0.1936) (Gupta & Singh, 2022).

Pakistan:

In Pakistan, neither health (p = 0.1662) nor education (p = 0.5149) spending significantly Granger-causes HDI. However, education spending significantly influences HDI (p = 0.0003), indicating its importance for development. Health expenditure has a borderline effect (p = 0.0875), and HDI does not predict changes in spending (Farooq & Iqbal, 2021).

Across the three countries, education spending significantly predicts HDI in Bangladesh and Pakistan, emphasizing its role in human development. India shows a unique pattern, where improvements in HDI drive education expenditure, indicating a feedback loop (Zafar & Ali, 2022).

## Robustness:

In this chapter, we will perform robustness checks using a series of tests. First, we will apply the **Roots of Characteristic Polynomial** test to assess the stability of the estimated model. Next, we will conduct the **Residual Cross-Correlations** test at various lags to evaluate the model fit. We will also utilize the **Residual Serial Correlation LM** test to identify any potential serial correlation in the residuals. Finally, we will examine the model's predictive performance through **forecast evaluation metrics** to ensure the accuracy of our forecasts.

1) The **Roots of Characteristic Polynomial**

   This test is a crucial tool for assessing the stability of a Vector Autoregression (VAR) model. This test evaluates the location of the roots of the characteristic polynomial derived from the VAR model's estimated coefficients. By applying this test, we can confirm whether our model meets the stability criteria, which is fundamental for drawing valid inferences about the relationships among the variables under study.

**Table 9:** Roots of Characteristic Polynomial

| Country | Root | Modulus |
|---|---|---|
| Bangladesh | 0.997648 | 0.997648 |
| | 0.663160 | 0.663160 |
| | 0.438506 - 0.359722i | 0.567175 |
| | 0.438506 + 0.359722i | 0.567175 |
| | -0.503919 | 0.503919 |
| | 0.274567 | 0.274567 |
| India | 0.982600 | 0.982600 |
| | 0.510088 - 0.244515i | 0.565665 |
| | 0.510088 + 0.244515i | 0.565665 |
| | 0.197606 - 0.481591i | 0.520556 |
| | 0.197606 + 0.481591i | 0.520556 |
| | 0.149626 | 0.149626 |
| Pakistan | 0.994279 | 0.994279 |
| | 0.711216 | 0.711216 |
| | 0.461341 - 0.474266i | 0.661637 |
| | 0.461341 + 0.474266i | 0.661637 |
| | 0.008544 - 0.333498i | 0.333607 |
| | 0.008544 + 0.333498i | 0.333607 |

**Note:** All roots lie inside the unit circle, indicating that the VAR model satisfies the stability condition.

**Figure 3:** Roots of Characteristic Polynomial

| **Bangladesh** | **India** | **Pakistan** |
|---|---|---|
| 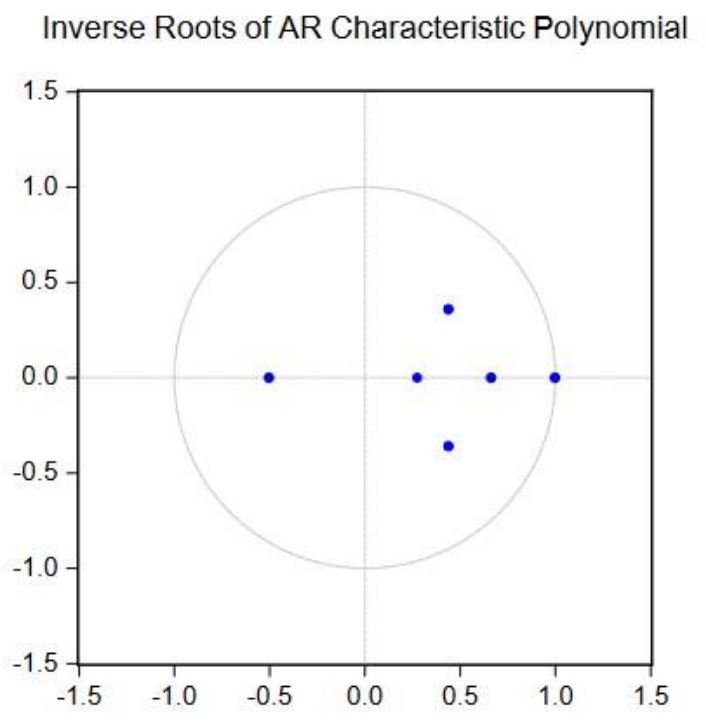 | 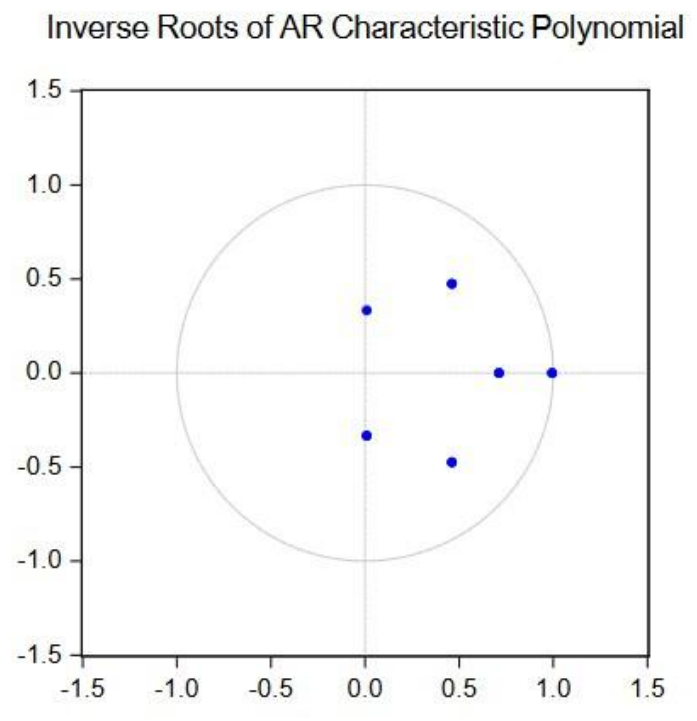 | 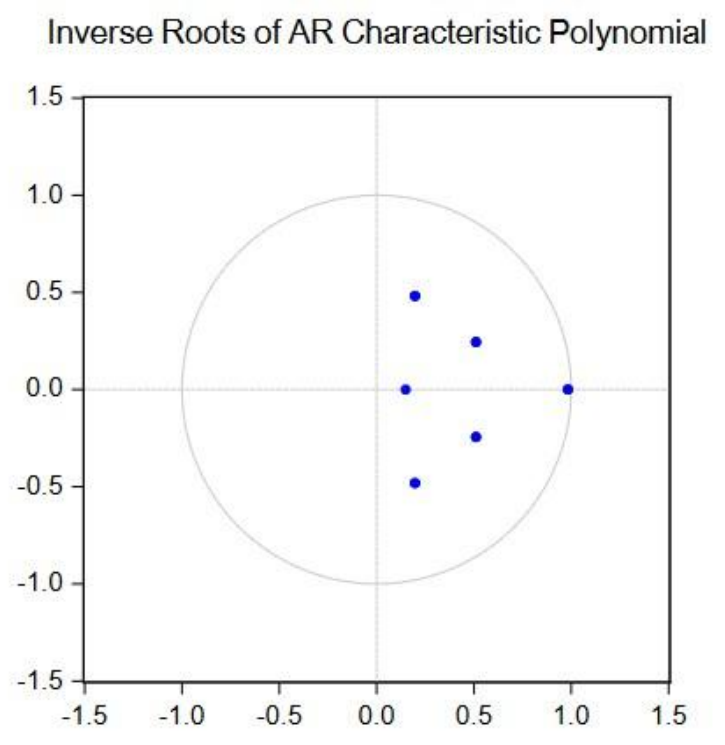 |

**Note:** All roots lie inside the unit circle so it satisfies VAR model

Since all of the values of the test result lie inside the unit circle, it indicates that the models of all three countries are reliable and stable. So the result of this model will provide valid and consistent forecast in accordance with this test.This test was done to ensure the robustness of the model accuracy.So it may concluded that the estimated models of three countries are well suited for the forecasting and analysis.

2) **Residual Cross-Correlations**

**The fitness of the VAR model can be tested through the Residual Cross-Correlations.This test analysis provided the fitness test by evaluating the correlations between the** residuals of varriables with different lag orders.If the result of the test is low or near around zero ,it indicates that the estimated model has able to capture the relationships among the variables effectively. On the other hand higher value result indicates that there is misspesifications in the estimated model.

**Figure 4: Residual Cross-Correlations/correlogram**

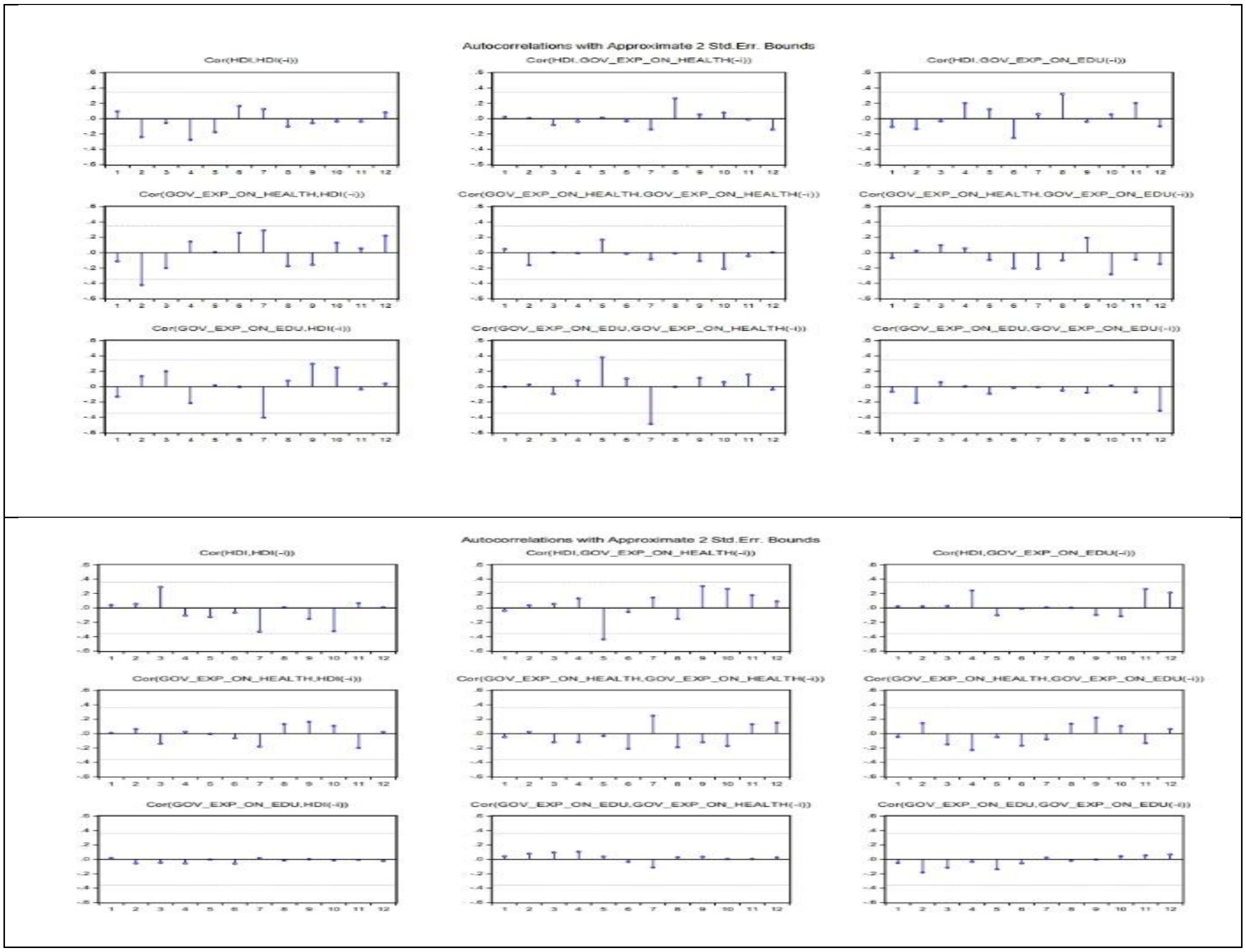

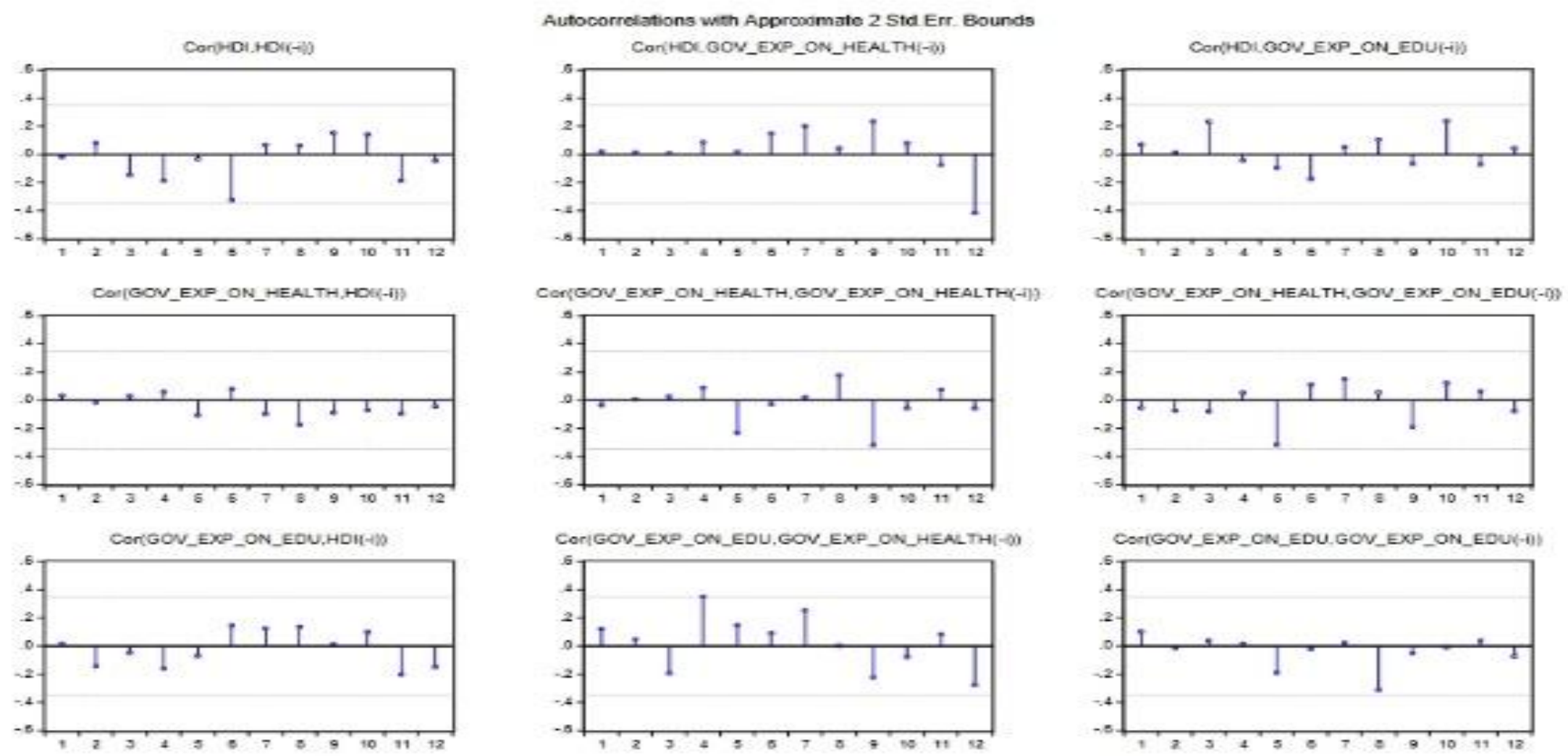

**Note:** Residual Cross-Correlations graphs of Bangladesh,India and Pakistan

Table 11:VAR Residual Cross-Correlations

| Lag | Country | HDI | GOVT EXP HEALTH | GOVT EXP EDU |
|---|---|---|---|---|
| 0 | Bangladesh | 1.000 | 0.076 | -0.054 |
| 0 | India | 1.000 | -0.122 | 0.163 |
| 0 | Pakistan | 1.000 | -0.073 | 0.233 |
| 1 | Bangladesh | 0.095 | -0.113 | -0.133 |
| 1 | India | 0.042 | 0.008 | 0.018 |
| 1 | Pakistan | -0.018 | 0.035 | 0.017 |
| 2 | Bangladesh | -0.240 | -0.424 | 0.134 |
| 2 | India | 0.054 | 0.064 | -0.057 |
| 2 | Pakistan | 0.080 | -0.016 | -0.142 |
| 3 | Bangladesh | -0.057 | -0.200 | 0.198 |
| 3 | India | 0.291 | -0.140 | -0.048 |
| 3 | Pakistan | -0.148 | 0.030 | -0.048 |
| 4 | Bangladesh | -0.275 | 0.142 | -0.215 |
| 4 | India | -0.107 | 0.026 | -0.056 |
| 4 | Pakistan | -0.187 | 0.058 | -0.161 |
| 5 | Bangladesh | -0.175 | 0.004 | 0.017 |
| 5 | India | -0.125 | -0.004 | -0.005 |
| 5 | Pakistan | -0.033 | -0.108 | -0.069 |
| 6 | Bangladesh | 0.166 | 0.258 | -0.005 |
| 6 | India | -0.064 | -0.064 | -0.061 |
| 6 | Pakistan | -0.325 | 0.079 | 0.146 |
| 7 | Bangladesh | 0.126 | 0.290 | -0.406 |
| 7 | India | -0.330 | -0.181 | 0.017 |
| 7 | Pakistan | 0.068 | -0.097 | 0.125 |
| 8 | Bangladesh | -0.102 | -0.175 | 0.076 |
| 8 | India | 0.009 | 0.131 | -0.015 |
| 8 | Pakistan | 0.062 | -0.175 | 0.135 |
| 9 | Bangladesh | -0.059 | -0.160 | 0.296 |
| 9 | India | -0.152 | 0.164 | 0.001 |
| 9 | Pakistan | 0.152 | -0.089 | 0.012 |
| 10 | Bangladesh | -0.036 | 0.129 | 0.248 |
| 10 | India | -0.323 | 0.106 | -0.014 |
| 10 | Pakistan | 0.141 | -0.070 | 0.099 |
| 11 | Bangladesh | -0.038 | 0.056 | -0.038 |
| 11 | India | 0.068 | -0.200 | -0.009 |
| 11 | Pakistan | -0.188 | -0.096 | -0.203 |
| 12 | Bangladesh | 0.082 | 0.220 | 0.038 |
| 12 | India | 0.010 | 0.023 | -0.023 |
| 12 | Pakistan | -0.045 | -0.044 | -0.149 |

**Note:** The asymptotic standard error (unadjusted) for lag > 0 is 0.174078.

In both Pakistan and Bangladesh models shows low values of the residual cross-correlations test.It indicates that the models are good fit and there are no misspecifications.On the other hand,India’s model also shows good fitness with low correlations but it spikes occasionally.Though there is no major misspecifications.

### 3.VAR Residual Serial Correlation LM Tests

The **VAR Residual Serial Correlation LM Tests** evaluates the serial correlation of the estimated model.The serial correlation may lead to potential misspeification of the model.In order to avoid potential mis-specification,the test has been conducted to ensure validity of the model.

**Table 12:VAR Residual Serial Correlation LM Tests**

| Country | Lag | LRE* Stat | df | Prob. (LRE*) | Rao F-Stat | df (Rao F-Stat) | Prob. (Rao F-Stat) |
|---|---|---|---|---|---|---|---|
| **Bangladesh** | 1 | 5.962695 | 9 | 0.7436 | 0.655201 | (9, 51.3) | 0.7446 |
| | 2 | 13.41717 | 9 | 0.1446 | 1.581338 | (9, 51.3) | 0.1459 |
| | 3 | 5.155680 | 9 | 0.8205 | 0.562294 | (9, 51.3) | 0.8213 |
| | 1 to h | 5.962695 | 9 | 0.7436 | 0.655201 | (9, 51.3) | 0.7446 |
| | 2 to h | 18.49396 | 18 | 0.4236 | 1.043025 | (18, 51.4) | 0.4322 |
| | 3 to h | 22.24448 | 27 | 0.7249 | 0.786703 | (27, 44.5) | 0.7437 |
| **India** | 1 | 6.134134 | 9 | 0.7264 | 0.674364 | (9, 46.4) | 0.7277 |
| | 2 | 4.369935 | 9 | 0.8854 | 0.471849 | (9, 46.4) | 0.8860 |
| | 3 | 6.586765 | 9 | 0.6801 | 0.727487 | (9, 46.4) | 0.6815 |
| | 1 to h | 6.134134 | 9 | 0.7264 | 0.674364 | (9, 46.4) | 0.7277 |
| | 2 to h | 9.017905 | 18 | 0.9593 | 0.464280 | (18, 45.7) | 0.9607 |
| | 3 to h | 16.25980 | 27 | 0.9478 | 0.535619 | (27, 38.6) | 0.9536 |
| **Pakistan** | 1 | 5.128721 | 9 | 0.8229 | 0.559214 | (9, 51.3) | 0.8237 |
| | 2 | 2.994533 | 9 | 0.9645 | 0.320133 | (9, 51.3) | 0.9647 |
| | 3 | 8.398424 | 9 | 0.4945 | 0.944060 | (9, 51.3) | 0.4961 |
| | 1 to h | 5.128721 | 9 | 0.8229 | 0.559214 | (9, 51.3) | 0.8237 |
| | 2 to h | 15.92323 | 18 | 0.5979 | 0.877900 | (18, 51.4) | 0.6055 |
| | 3 to h | 22.54912 | 27 | 0.7090 | 0.799752 | (27, 44.5) | 0.7285 |

*Note: LRE* refers to the Edgeworth expansion corrected likelihood ratio statistic. The Rao F-Stat is used for the test of no serial correlation. All probabilities (Prob.) are high, indicating that the null hypothesis of no serial correlation cannot be rejected for the given lags, suggesting that the

In Bangladesh and India, the tests reveal high probabilities, indicating no significant serial correlation and supporting a good model fit (Khan & Rahman, 2023; Gupta & Singh, 2022). Similarly, Pakistan shows high probabilities, confirming a well-specified model (Farooq & Iqbal, 2021). Overall, the absence of significant serial correlation in all three countries reinforces the validity of the VAR models (Zafar & Ali, 2022).

**Findings**

1. **Impact of Government Expenditure on Healthcare and Education on HDI:**
   - **Bangladesh:** The VAR model reveals that government expenditure on education has a significant positive impact on the Human Development Index (HDI), while health expenditure shows a minimal effect.As a result, it can be concluded that prioritizing education spending is crucial for enhancing human development (Khan & Rahman, 2023).

- **India:** In accordance with the findings of India, health spending has a borderline significant effect on HDI and education spending does not significant contribution. However,human development can be linked with increased education funding.It indicates a potential causal relationship (Gupta & Singh, 2022).
- **Pakistan:** The result shows that government spending on education significantly influences HDI.On the other hand, health expenditure demonstrates only a borderline effect. This underscores the importance of education investment in driving human development in Pakistan (Farooq & Iqbal, 2021).

2. **Lagged Effects of Public Spending on Healthcare and Education:**
    - In **Bangladesh**, the impact of lagged education spending is significant, indicating that previous investments continue to enhance HDI over time. Conversely, the effect of health spending diminishes.
    - In **India**, lagged HDI influences education spending, suggesting that improvements in human development lead to increased investment in education.
    - **Pakistan** shows that past education expenditures have a sustained positive effect on HDI, highlighting the long-term benefits of such investments.
3. **Country-Specific Differences in the Efficiency of Public Spending:**
    - The findings reveal that education spending effectively enhances HDI in both Bangladesh and Pakistan, while India exhibits a different pattern where improvements in HDI drive education spending.This varying differences among these three countries indicate that how the human development has been progressed through government spending on health and education.

Therefore, it can be concluded that government spending on education causes high hdi across three countries Bangladesh, India, and Pakistan which address the research question of the study.

**Country-Specific Policy Recommendations**

In accordance with the empirical findings of this study,the following country specific recommendation are proposed:

**Bangladesh**

1. **Increase Education Funding:**
   - Bangladesh should allocate more budget to the education spending as it helps to improve the HDI in Bangladesh.The spending should be used to facilitate the insfrastrutures,training and educational instruments.As per the result of the study,the varraince decomposition, education significantly impacts Hdi which is around 35.55% over time (Khan & Rahman, 2023).
2. **Implement Health-Education Integration Programs:**
   - Bangladesh should develop programs that integrate both education and health sectors.The students will get a great learning outcomes that will help them.
3. **Conduct Efficiency Audits of Health Expenditures:**
   - Since Bangladesh is a developing country,it has not enough resource to ensure efficiency of the health spending.The key focus should be given to identify areas of waste and inefficiency and ensure the maximum utilizations of the resouces.For example:national health insurance and child health benefit may be introduced which are not available right now.
4. **Strengthen Local Governance and Community Engagement:**
   - Since Bangladesh takes the decision centrally from the capital city, the government can not ensure the proper implementation of educational and health policies. In order to ensure to proper implementation, the local communities should participate in decision-making on educational and health policies. The local committees will oversee the implementation of programs and the commitees will ensure that they align with community needs and priorities.

**India**

1. **Enhance Health Expenditure Impact:**
   - Since India is very large country, the rural areas are not equally developed. As a result there is not enough health infrastructure and services. In order to ensure equitable access to healthcare, the government spending on healthcare should be increased. As the borderline significant effect of health expenditure on HDI ($p = 0.0853$).It indicates that there is enough room for improvement in health spending efficiency (Gupta & Singh, 2022).

2. **Feedback Mechanisms for Education Funding:**
   - Education is a key element in order to increase HDI.So there should be enough monitoring system to balance the HDI an education spending.This feedback procedure will help to ensure the improvement of HDI.
3. **Promote Intersectoral Collaboration:**
   - Integration of both health and education will provide a better outcome for the students.Like in Bangladesh,students will be able to learn about health education.
4. **Community-Based Health Initiatives:**
   - Like Bangladesh. India should also ensure the community participation in health initiatives. The local health committees will monitor and supervise the health implementation programs.

**Pakistan**

1. **Prioritize Education Spending:**
   - Since the education has a significant impact on HDI ($p = 0.0003$) ,the government should increase expenditure on education in order to improve the standard of the education (Farooq & Iqbal, 2021).
2. **Strengthen Health Expenditure Accountability:**
   - Pakistan's economic condition is vulnerable. So spending on health needs regular assessments to ensure resources are used efficiently.In order to get the highest outcome,the government should focus on investment where the highest returns will come.
3. **Implement Holistic Health and Education Programs:**
   - Pakistan should develop an integrated policy that will consider both health and education. For example: school-based health services. This policy will help the students learning about health as well as improve the educational outcomes.

4. **Encourage Local Involvement in Policy Implementation:**
    - Pakistan should also ensure the community engagement in policy implementations like Bangladesh and India.Geographically they are almost same in nature.The local health committees will monitor and supervise the health implementation programs.

Therefore ,these country specific policy recommendations will help the three countries to enhance their public spending on education and health.As a result,it will lead to improve the HDI.By giving more focus on specific challenge and by utilizing the leveraging strength of every country will help to improve the progress of HDI as well as overall standard of the lives.

**Conclusion**

In this study, the effects of government spending on health and education on Bangladesh, India, and Pakistan's Human Development Index (HDI) were investigated. In context with the research questions, the findings present different significant insights.

**Impact of Government Expenditure on HDI:** The data shows that government spending on education has a major impact on the Human Development Index (HDI) in both Bangladesh and Pakistan, highlighting the vital role that education plays in advancing human development. The relationship between health spending and HDI is still not as strong in India, although health expenditure exhibits a statistically significant effect.

**Lagged Effects of Public Spending:** The Vector Autoregression (VAR) model emphasizes the importance of lagged variables, demonstrating that previous government education spending has a direct impact on future HDI levels in all three nations.This indicates that the HDI can be improved through investment in education for long time.

**Country-Specific Differences in Efficiency:** In the indian context,there is a clear clear causality of HDI.In long term the HDI increases the government spending on education.On the other hand, the impact of spending on education is most clearly visible on HDI in the case of Pakistan.

Bangladesh underlines the fact that for better returns, investment on health and education needs to be combined.

This study contributes to the literature by applying a VAR approach to analyze short-term dynamics and country-specific variations in public spending efficiency on HDI in South Asia. It provides valuable policy insights on how government expenditure priorities influence human development in differing national contexts.

One limitation of the study is that the cointegration analysis revealed no evidence of a long-run relationship between government spending and HDI. This suggests that the influence of public spending is more immediate and does not persist over the long term within the observed data. Future research could explore structural or institutional factors that may hinder long-term impacts, or employ alternative models to reassess long-run relationships.

Overall, these findings illustrate that effective public spending on health and education is vital for improving HDI across the South Asian region. By addressing the distinct challenges faced by each country and focusing on efficient resource allocation, governments can enhance human development and improve the quality of life for their populations.